\documentstyle[12pt]{article}
\newcommand{\lambdaslash}{\not \! \lambda}

\begin{document}

\begin{flushright}
%July 19, 2008
\end{flushright}

\vskip 0.5 truecm

\begin{center}
{\Large{\bf Clear evasion of the uncertainty relation with very small probability}}
\end{center}
\vskip .5 truecm
\centerline{\bf  Kazuo Fujikawa and Koichiro Umetsu}
\vskip .4 truecm
\centerline {\it Institute of Quantum Science, College of 
Science and Technology}
\centerline {\it Nihon University, Chiyoda-ku, Tokyo 101-8308, 
Japan}
\vskip 0.5 truecm

\makeatletter
\@addtoreset{equation}{section}
\def\theequation{\thesection.\arabic{equation}}
\makeatother

\begin{abstract}
We entertain the idea that the uncertainty relation is not a
principle, but rather it is a consequence of quantum mechanics.
The uncertainty relation is then a probabilistic statement and 
can be clearly evaded in processes which occur with a very small 
probability in a tiny sector of the phase space. This clear 
evasion is typically realized when one utilizes indirect 
measurements, and some examples of the clear evasion appear
in the system with entanglement though the entanglement by 
itself is not essential for the evasion.  The standard Kennard's relation and its interpretation remain intact in our analysis. 
 As an explicit example, we show that the clear 
evasion of the uncertainty relation for coordinate and momentum 
in the diffraction process discussed by
Ballentine is  realized in a tiny sector  of the 
phase space with a very small probability. 
We also examine the uncertainty relation for a two-spin system with the EPR entanglement and show that no clear evasion takes place in this system with the finite discrete degrees of freedom.
\end{abstract}

%\large

\section{Introduction}

The Schr\"{o}dinger amplitude together with the  Schr\"{o}dinger
equation describes all the possible quantum states and their 
time developments. Combined with the Born probability 
interpretation we have the rules to interpret quantum 
mechanics~\cite{dirac}.
The Heisenberg uncertainty relation~\cite{heisenberg} (and its 
reformulation in~\cite{kennard, robertson}) and the EPR entanglement~\cite{epr}
represent two characteristic features of quantum mechanics
which are foreign to macroscopic phenomena in classical physics. 
The ultimate purpose of the 
theory of measurements such as described in~\cite{neumann,margenau, 
arthurs, ballentine, braginsky, appleby, appleby2,  busch, 
hilgevoord, uffink,busch2} is to explain how to understand 
the observed phenomena starting with the basic rules of 
quantum theory.  
  
It is customary to treat the uncertainty relation as a 
principle, namely, the {\em uncertainty principle} which defines 
the quantum mechanics at the deepest level. In this paper, we 
entertain the idea that the uncertainty relation is one of the 
consequences of quantum mechanics. To be more precise, the 
uncertainty relation is controlled by the probability 
interpretation of quantum mechanics, and thus the validity 
of the uncertainty relation is probabilistic one. If one 
allows a very small probability in a tiny sector of the phase
space, a clear evasion of the uncertainty relation is allowed.

We illustrate this idea by using the clear evasion of the 
uncertainty relation in the diffraction process discussed by 
Ballentine~\cite{ballentine} some time ago. We also discuss two
 simple gedanken experiments considered by 
Ozawa~\cite{ozawa-pl03, ozawa-pl04} as examples of the clear evasion of the uncertainty relation in the present sense.
The evasion is typically recognized  when one utilizes
indirect measurements, and some examples of the clear evasion appear
in the system with entanglement though the entanglement by 
itself is not essential for the evasion. 
The standard Kennard's relation~\cite{kennard} is naturally preserved intact in our analysis.

It should be emphasized that the evasion of the uncertainty
relation we discuss has no connection with the
analysis of the differences between the uncertainty relation 
in the sense of Kennard~\cite{kennard} and the uncertainty 
relation appearing in the detailed definition of measurements. 
The evasion of the uncertainty relation in this latter sense 
mainly analyzes the model of quantum  measurements initiated by 
von Neumann~\cite{neumann} and its modifications. This class of 
violation does not associate the violation of the uncertainty 
relation with a very small probability in a tiny sector of the 
phase space and often discusses the evasion of the quantum 
limit of measurements. The notion such as the {\em evasion of 
the 
quantum limit} does not arise in our analysis, since we operate
in the framework of quantum mechanics, and the occurrence of 
the violation takes place simply with a very small probability. 
Our analysis is thus perfectly consistent with 
experimental results so far that no clear evasion of the 
uncertainty relation has been reported.

Our analysis mainly utilizes indirect measurements, as we noted 
above, and the typical indirect measurement is realized for a 
system with  entanglement. We thus briefly comment on the 
uncertainty relation  for the two-spin system with the EPR 
entanglement~\cite{epr}. We
show that the clear evasion of the uncertainty relation is not 
realized for the two-spin system with the finite discrete 
degrees of freedom, since a tiny sector of the phase space with 
a very small probability is not simply defined for a finite 
discrete system.

\section{Indirect measurement and uncertainty relation}

We here discuss the clear evasion of the uncertainty relation in several simple 
processes by utilizing indirect measurements. Our discussion 
heavily relies on the explicit model of  
Ballentine~\cite{ballentine} for a diffraction of a 
particle by a single slit (or a pin hole), which has been 
discussed without an emphasis on a very small probability.

\subsection{Diffraction and uncertainty relation}

Consider the plane wave of a particle with momentum $p$ moving 
toward the positive x-direction and colliding with a plate which
 is placed perpendicular to the x-axis at $x=0$ with a pin hole 
at the origin of the y-z plane. See Fig. 1. The 
particle diffracted in the forward direction is measured on the 
screen which is perpendicular to the x-axis and placed at $x=L$.

\normalsize
\begin{center}
%WinTpicVersion3.08
\unitlength 0.1in
\begin{picture}( 32.0000, 20.0000)(  2.0000,-24.0000)
% BOX 2 0 3 0
% 2 1000 400 1060 1330
% 
\special{pn 8}%
\special{pa 1000 400}%
\special{pa 1060 400}%
\special{pa 1060 1330}%
\special{pa 1000 1330}%
\special{pa 1000 400}%
\special{fp}%
% VECTOR 2 0 3 0
% 2 200 1400 600 1400
% 
\special{pn 8}%
\special{pa 200 1400}%
\special{pa 600 1400}%
\special{fp}%
\special{sh 1}%
\special{pa 600 1400}%
\special{pa 534 1380}%
\special{pa 548 1400}%
\special{pa 534 1420}%
\special{pa 600 1400}%
\special{fp}%
% VECTOR 2 0 3 0
% 2 790 1400 3400 1400
% 
\special{pn 8}%
\special{pa 790 1400}%
\special{pa 3400 1400}%
\special{fp}%
\special{sh 1}%
\special{pa 3400 1400}%
\special{pa 3334 1380}%
\special{pa 3348 1400}%
\special{pa 3334 1420}%
\special{pa 3400 1400}%
\special{fp}%
% BOX 2 0 3 0
% 2 1000 1470 1060 2400
% 
\special{pn 8}%
\special{pa 1000 1470}%
\special{pa 1060 1470}%
\special{pa 1060 2400}%
\special{pa 1000 2400}%
\special{pa 1000 1470}%
\special{fp}%
% STR 2 0 3 0
% 3 810 1500 810 1600 0 0
% 
\put(8.1000,-16.0000){\makebox(0,0)[lb]{}}%
% STR 2 0 3 0
% 3 940 1380 940 1480 4 0
% $0$
\put(9.4000,-14.8000){\makebox(0,0)[rt]{$0$}}%
% LINE 2 0 3 0
% 2 2800 410 2800 2400
% 
\special{pn 8}%
\special{pa 2800 410}%
\special{pa 2800 2400}%
\special{fp}%
% STR 2 0 3 0
% 3 3490 1300 3490 1400 5 0
% $x$
\put(34.9000,-14.0000){\makebox(0,0){$x$}}%
% VECTOR 2 0 3 0
% 4 1800 1600 1060 1600 2000 1600 2790 1600
% 
\special{pn 8}%
\special{pa 1800 1600}%
\special{pa 1060 1600}%
\special{fp}%
\special{sh 1}%
\special{pa 1060 1600}%
\special{pa 1128 1620}%
\special{pa 1114 1600}%
\special{pa 1128 1580}%
\special{pa 1060 1600}%
\special{fp}%
\special{pa 2000 1600}%
\special{pa 2790 1600}%
\special{fp}%
\special{sh 1}%
\special{pa 2790 1600}%
\special{pa 2724 1580}%
\special{pa 2738 1600}%
\special{pa 2724 1620}%
\special{pa 2790 1600}%
\special{fp}%
% STR 2 0 3 0
% 3 1900 1500 1900 1600 5 0
% $L$
\put(19.0000,-16.0000){\makebox(0,0){$L$}}%
% VECTOR 2 0 3 0
% 2 1030 1400 2800 630
% 
\special{pn 8}%
\special{pa 1030 1400}%
\special{pa 2800 630}%
\special{fp}%
\special{sh 1}%
\special{pa 2800 630}%
\special{pa 2732 638}%
\special{pa 2752 652}%
\special{pa 2748 676}%
\special{pa 2800 630}%
\special{fp}%
% STR 2 0 3 0
% 3 400 1430 400 1530 5 0
% $p$
\put(4.0000,-15.3000){\makebox(0,0){$p$}}%
% VECTOR 2 0 3 0
% 4 2925 530 2925 730 2925 730 2925 530
% 
\special{pn 8}%
\special{pa 2926 530}%
\special{pa 2926 730}%
\special{fp}%
\special{sh 1}%
\special{pa 2926 730}%
\special{pa 2946 664}%
\special{pa 2926 678}%
\special{pa 2906 664}%
\special{pa 2926 730}%
\special{fp}%
\special{pa 2926 730}%
\special{pa 2926 530}%
\special{fp}%
\special{sh 1}%
\special{pa 2926 530}%
\special{pa 2906 598}%
\special{pa 2926 584}%
\special{pa 2946 598}%
\special{pa 2926 530}%
\special{fp}%
% STR 2 0 3 0
% 3 2965 600 2965 700 2 0
% $\delta y$
\put(29.6500,-7.0000){\makebox(0,0)[lb]{$\delta y$}}%
% STR 2 0 3 0
% 3 2950 910 2950 1010 5 0
% $y$
\put(29.5000,-10.1000){\makebox(0,0){$y$}}%
% VECTOR 2 0 3 0
% 4 2840 610 2840 1410 2840 1410 2840 610
% 
\special{pn 8}%
\special{pa 2840 610}%
\special{pa 2840 1410}%
\special{fp}%
\special{sh 1}%
\special{pa 2840 1410}%
\special{pa 2860 1344}%
\special{pa 2840 1358}%
\special{pa 2820 1344}%
\special{pa 2840 1410}%
\special{fp}%
\special{pa 2840 1410}%
\special{pa 2840 610}%
\special{fp}%
\special{sh 1}%
\special{pa 2840 610}%
\special{pa 2820 678}%
\special{pa 2840 664}%
\special{pa 2860 678}%
\special{pa 2840 610}%
\special{fp}%
\end{picture}%

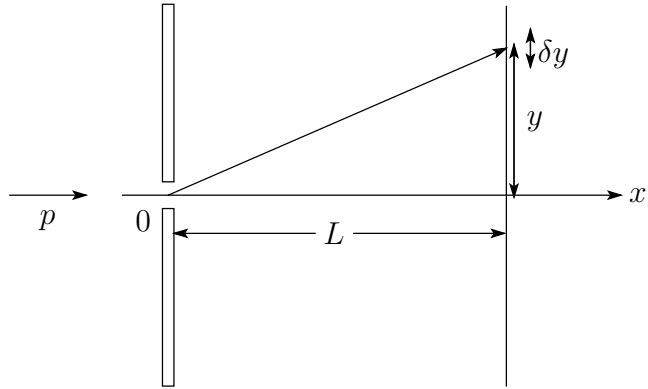
Fig.1. A schematic arrangement to measure $y$ and then infer the 
value of $p_{y}$.
\end{center}

%\large
Suppose that one detects the diffracted particle by a detector 
with a size $\delta y$ which is placed on the y-axis of the 
screen at the position $y$.   
The y-component of the momentum of the measured particle is 
indirectly estimated  at
\begin{eqnarray}
p_{y}=p\frac{y}{\sqrt{L^{2}+y^{2}}}
\end{eqnarray}
and the coordinate accuracy is $\delta y$. The uncertainty in
$p_{y}$ is estimated by 
\begin{eqnarray}
\delta p_{y}=p\frac{\delta y}{\sqrt{L^{2}+y^{2}}}
\frac{L^{2}}{L^{2}+y^{2}}\sim p 
\frac{\delta y}{L}
\end{eqnarray}
for fixed $|y|$ with $L\geq |y|$.
The uncertainty product is then
\begin{eqnarray}
\delta y\delta p_{y}\sim p\delta y \frac{\delta y}{L}
\rightarrow 0
\end{eqnarray} 
for
$L\rightarrow\infty$ for fixed $\delta y$. 
Thus the conventional uncertainty relation appears to be  
clearly evaded. See Subsection 3.2 in~\cite{ballentine}.

In the idealized analysis of Ballentine~\cite{ballentine}, no 
variable $\delta l$ corresponding to the {\em size of the 
opening} of the slit appears. A slit of $\delta l$ induces a 
disturbance
\begin{eqnarray}
\delta p_{y}\sim \hbar/\delta l
\end{eqnarray}
in the momentum of the particle immediately after the passage of
 the slit and it causes the effect of diffraction, but the above
 conclusion
is still valid as long as the scattering by the slit is elastic
and thus the energy conservation of the combined system of 
the particle and the slit holds, namely, the magnitude of the 
momentum of the 
particle after the passage of the slit is still given by  $p$
to a good accuracy. To be precise, one may choose 
\begin{eqnarray}
p\geq \hbar/\delta l
\end{eqnarray}
or
\begin{eqnarray}
\delta l\geq \hbar/p= \lambdaslash 
\end{eqnarray}
but not much larger than $\lambdaslash$, where $\lambdaslash$
stands for the de Broglie wave length of the particle. Our assumption is that the slit with a large mass absorbs the momentum but no energy is lost due to the possible excitation of the atomic states in the slit, as in the case of the M\"{o}ssbauer effect.

The setup of the problem in~\cite{ballentine}
is then analogous to the elastic scattering of a particle by a
heavy target located at the center of the slit. To define the 
geometrical picture in Fig.1 precisely, one needs to set 
$L\rightarrow\infty$. In this case, one may rescale all 
the length variables by $L$ as in the scattering problem, and 
thus 
\begin{eqnarray}
\tilde{y}=\frac{y}{L},\ \ \ \ \ \delta \tilde{y}
=\frac{\delta y}{L}
\end{eqnarray}
and one may describe the scattering in terms of $\tilde{y}$ and 
$\delta \tilde{y}$. The uncertainty product (2.3) is then given by 
\begin{eqnarray}
\delta y\delta p_{y}\sim p\delta y \frac{\delta y}{L}
\simeq \hbar \frac{L}{\lambdaslash}(\delta \tilde{y})^{2}\gg
\hbar
\end{eqnarray} 
for fixed finite $\delta \tilde{y}$, where $\lambdaslash$ stands for the de Broglie wave length, and the issue of the uncertainty relation does not arise  in this description. This description may be regarded as corresponding to a  description in the classical limit.

The clear evasion of the uncertainty relation (2.3)
was possible since we restricted the range $\delta y$ to a 
measure-zero set in the {\em a priori} allowed interval $(0, \infty)$, namely, 
$\delta y/\tilde{L}\rightarrow 0$ for 
$L\rightarrow {\rm large}$ where $\tilde{L}$ stands for the size of the spread of the diffracted particle at $x=L$.   
A way to understand the special property of (2.3) is 
to consider the wave function of the diffracted particle which may be represented by
\begin{eqnarray} 
\psi(x,y,z)= \frac{1}{r}\exp[ipr/\hbar]f(\theta)
\end{eqnarray}
by assuming the symmetry around the x-axis. Here $r=\sqrt{x^{2}+\rho^{2}}$, $\rho=\sqrt{y^{2}+z^{2}}$ and $\sin\theta=\rho/r$;
$\psi(x,y,z)$ may be normalized to unity when integrated over a large hemi-shepre with radius $r$ since $x\geq 0$ in Fig.1. One may then imagine to detect the particle which arrives at a ring (annulus) specified by the radius $\rho$ and 
$\rho+\delta\rho$ on the screen placed at $x=L$.
The probability to detect the particle is then 
\begin{eqnarray}
 |\psi(x,y,L)|^{2}2\pi\rho\delta\rho=
 |\frac{1}{r}f(\theta)|^{2}2\pi\rho\delta\rho\sim
\frac{\rho\delta\rho}{r^{2}}\leq\frac{\delta\rho}{r}\rightarrow 0
\end{eqnarray}
for $L\rightarrow\infty$, and thus the probability is vanishingly small; $\delta\rho$ here corresponds to $\delta y$, and this conclusion also holds when one considers the flux instead of the probability. 
The clear evasion of the uncertainty relation for the set of events with a very small probability 
does not contradict the principle of quantum mechanics. In the conventional treatment of scattering, one fixes $\delta\sin\theta
=\delta\rho/r$ for $L\rightarrow\infty$ and no evasion of the uncertainty relation occurs. See (2.8).

Immediately {\em after} the 
measurement on the screen, however, the particle is described by
 a wave packet of the size $\delta \rho$ in the $\rho$ direction.
 Translated into the notation of $y$, the wave function $\psi(y)$ (by suppressing the dependence on other variables) does not contain the small factor  $\sim 1/\sqrt{\tilde{L}}$ in front of it (as in the case of approximate plane wave $\sim (1/\sqrt{\tilde{L}})\exp[ip_{y}y/\hbar]$ before the detection) due to the ``reduction'' of the wave function.  At the same time, we generally expect $\delta p_{y}\sim \hbar/\delta y$ after the 
measurement. The event realized with nearly a unit probability such as the present example after the actual measurement does not lead to a clear evasion of the uncertainty relation. More comments on this problem will be given later.

\subsection{Other examples of the clear evasion of the uncertainty relation}

We next study a two-particle system in one-dimensional space 
with an equal mass $m$, and their phase space variables are 
$(x_{1}, p_{1})$ and $(x_{2}, p_{2})$. We then define
\begin{eqnarray}
&&P=p_{1}+p_{2}, \ \ Q=\frac{1}{2}(x_{1}+x_{2}),\nonumber\\
&&p=\frac{1}{2}(p_{1}-p_{2}), \ \ q=x_{1}-x_{2}.
\end{eqnarray}

An example given by Ozawa 
 for which the clear evasion of 
the uncertainty relation takes place  
goes as follows (see Section 9 in~\cite{ozawa-pl03}): Starting with a two-particle system such as  
(2.11), one assumes the precise measurement of $x_{2}$ and no 
more measurements  of other variables. Thus the measurement is 
rather incomplete for the system 
of two degrees of freedom. Apparently the disturbance 
$\eta(p_{1})$ in 
the momentum $p_{1}$ of the first particle, which is defined in 
Appendix, caused by the 
measurement of $x_{2}$ vanishes 
\begin{eqnarray}
\eta(p_{1})=0.
\end{eqnarray}
It is shown that the noise (or error) in the coordinate of
the first particle $x_{1}$, when the precise measurement of 
$x_{2}$ is regarded as an {\em indirect} measurement of 
$x_{1}$, is given by 
$\epsilon(x_{1})=\langle (\hat{x}_{2}-\hat{x}_{1})^{2}\rangle^{1/2}$
for any input state~\footnote{
The precise measurement of $x_{2}$ implies
$\hat{\mu}_{X_{2}f}=\hat{x_{2}}$ where $\hat{\mu}_{X_{2}f}$ stands for the 
apparatus variable. The noise in $\hat{x}_{1}$ when one regards
 $\hat{\mu}_{X_{2}f}$ as the measuring apparatus of $\hat{x}_{1}$ is 
then given by $\epsilon (x_{1})=\langle (\hat{\mu}_{X_{2}f} -
\hat{x}_{1})^{2}\rangle^{1/2}
=\langle(\hat{x_{2}}-\hat{x}_{1})^{2}\rangle^{1/2}$. 
See Section 9 in~\cite{ozawa-pl03}.}. Note that 
the states in the present paper stand
 for initial states  as we work in the Heisenberg picture, and 
the dynamical variables in our examples always stand for initial
 variables. Ozawa then argues that one can choose  input initial states which satisfy
\begin{eqnarray}
\epsilon(x_{1})=\langle (\hat{x}_{2}-\hat{x}_{1})^{2}
\rangle^{1/2}<\alpha
\end{eqnarray}
for any 
small $\alpha$. He thus concludes~\cite{ozawa-pl03} that one can
estimate $x_{1}$ with $\epsilon(x_{1})<\alpha$ for any 
small $\alpha$ from the measured
 value of $x_{2}$  without disturbing $p_{1}$, namely, $\eta(p_{1})=0$ for all such states, and thus the clear evasion of the uncertainty relation $\epsilon(x_{1})\eta(p_{1})\geq \frac{1}{2}\hbar$ in the sense of (A.8) in Appendix. 

Without any information  about 
the total system (namely, without any measurement of $x_{1}$), 
the {\em a priori} natural guess is
\begin{eqnarray}
\langle (\hat{x}_{2}-\hat{x}_{1})^{2}\rangle^{1/2}\sim L
\end{eqnarray}
where $L$ stands for the size of the ``box'' in which the system
 is contained. In this sense
\begin{eqnarray}
\Delta(\hat{x}_{2}-\hat{x}_{1})\leq 
\langle (\hat{x}_{2}-\hat{x}_{1})^{2}\rangle^{1/2}<\alpha
\end{eqnarray}
for any small $\alpha$, where $\Delta(\hat{x}_{2}-\hat{x}_{1})$
stands for the standard deviation, implies a significant correlation.
Under this assumption of significant correlation, one can 
estimate the initial value 
of $x_{1}$  without disturbing $p_{1}$ by using the precisely 
measured value of $x_{2}$. 
Stated differently,
one can know $x_{1}$ accurately without disturbing $p_{1}$ by 
measuring $x_{2}$ precisely, if the 
particle 1 happens to be very close to the particle 2.

The above argument related to (2.13) shows that one 
can clearly evade the uncertainty relation 
for any input wave function which is confined in a 
tiny subspace $\langle (\hat{x}_{2}-\hat{x}_{1})^{2}
\rangle^{1/2}<\alpha$ of the allowed range of the variable $x_{1}$ with a precisely measured $x_{2}$.
This subspace is measure-zero compared to the full one-dimensional space, namely,
 the interval covered by the eigenvalues with 
$\langle (\hat{x}_{2}-\hat{x}_{1})^{2}\rangle^{1/2}<\alpha$ is negligible 
compared to the total space of {\em a priori} allowed 
eigenvalues  $-L<x_{1}-x_{2}<L$ for $L\rightarrow {\rm large}$ 
 with a fixed $x_{2}$. The probability of observing such an event is 
 very small. 
 
This small probability is understood by considering a generic normalized wave function $\Psi(x_{1},x_{2})\simeq \sum_{n} a_{n}\psi_{n}(x_{2},q)$ 
where $\psi_{n}(x_{2},q)$ stand for normalized non-overlapping 
wave packets in $q$ with a width $\alpha$. The function 
 $\psi_{n}(x_{2},q)$ may have a peak around the precisely measured value of $x_{2}$ and the coefficients are 
$|a_{n}|\sim 1/\sqrt{2L/\alpha}$ without any additional information about $q$. Here we have a Bloch-type representation of $\Psi(x_{1},x_{2})$  in mind. The above specific choice of the initial state in (2.13) corresponds to selectively picking up a specific  $\psi_{n}(x_{2},q)$ which contains $q=x_{1}-x_{2}=0$, but the standard interpretation of the wave function $\Psi(x_{1},x_{2})$ in  quantum mechanics predicts that such a probability is negligibly small 
\begin{eqnarray}
|a_{n}|^{2}\sim \alpha/(2L)
\end{eqnarray}
 for $L\rightarrow {\rm large}$. 
 In contrast, after the actual direct measurement of the state $\psi_{n}(x_{2},q)$ which contains $q=x_{1}-x_{2}=0$, such a state is then realized with a unit probability due to the reduction of the state 
 vector. The momentum disturbance  $\eta(p_{1})$ is then generally modified 
 as predicted by quantum mechanics for the state $\psi_{n}(x_{2},q)$ and thus no clear evasion of the uncertainty relation. (To be precise,
the condition $\eta(p_{1})=0$ may be still preserved even after the 
actual measurement of the state $\psi_{n}(x_{2},q)$ which contains $q=x_{1}-x_{2}=0$, but such a probability is negligibly small as is understood by writing $\psi_{n}(x_{2},q)$ as a superposition of plane waves. The probability of a clear evasion of the uncertainty relation is very small in any case.) 
 
 In passing, the average $\langle (\hat{x}_{2}-\hat{x}_{1})^{2}\rangle$ for the state $\Psi(x_{1},x_{2})$ is estimated by  
\begin{eqnarray}
\langle (\hat{x}_{2}-\hat{x}_{1})^{2}\rangle\sim\sum_{q}
q^{2}\frac{\alpha}{2L}\sim \int^{L}_{-L}dq q^{2}\frac{1}{2L} \sim \frac{L^{2}}{3}\nonumber
\end{eqnarray}
and thus $\langle (\hat{x}_{2}-\hat{x}_{1})^{2}\rangle^{1/2}\sim L$ which is consistent with (2.14) and does not give rise to a clear evasion of the uncertainty relation.

Another example of the clear evasion of the uncertainty relation 
given by Ozawa~\cite{ozawa-pl04} goes as follows:
When one measures the commuting variables $\hat{P}$ and 
$\hat{q}$  precisely with $\epsilon(P)=\epsilon(q)=0$,
it is then argued that
\begin{eqnarray}
\epsilon(x_{1})\epsilon(P)=0
\end{eqnarray}
and thus the clear evasion of the uncertainty relation 
in the sense of (A.7) in Appendix for the pair $[\hat{x}_{1},\hat{P}]=i\hbar$.
Here $\epsilon(x_{1})$ and $\epsilon(P)$ respectively stand for
 the operationally defined root-mean-square noise in $x_{1}$ and 
$P$ which are briefly 
explained in Appendix.     

The essence of 
the  argument for the above relation (2.17) is as 
follows(see Section 2 in~\cite{ozawa-pl04}): The precise 
measurement of $q$ can be interpreted as an approximate 
{\em indirect} measurement 
of $x_{1}$, if one takes the output from the measurement of $q$ 
to be 
the measured value of $x_{1}$. 
The root-mean-square noise~\footnote{
The precise measurement of $q$ implies
$\hat{\mu}_{Qf}=\hat{q}$ where $\hat{\mu}_{Qf}$ stands for the 
apparatus variable. The noise in $\hat{x}_{1}$ when one 
regards $\hat{\mu}_{Qf}$ as the measuring apparatus of 
$\hat{x}_{1}$ is 
then given by $\epsilon (x_{1})=\langle (\hat{\mu}_{Qf} -
\hat{x}_{1})^{2}\rangle^{1/2}
=\langle(\hat{q}-\hat{x}_{1})^{2}\rangle^{1/2}$. } 
\begin{eqnarray}
\epsilon (x_{1})=\langle (\hat{q}-\hat{x}_{1})^{2}\rangle^{1/2}=
\langle (\hat{x}_{2})^{2}\rangle^{1/2}
\end{eqnarray}
is thus a reasonable measure of imprecision for the $x_{1}$ 
measurement. 

If the mean position of $x_{2}$ is chosen at the origin, 
$\langle \hat{x}_{2}\rangle=0$, then one has 
$\epsilon (x_{1})=\Delta x_{2}$.
Ozawa then argues that $\epsilon (x_{1})=\Delta x_{2}$ can be 
made arbitrarily small 
\begin{eqnarray}
\epsilon (x_{1})=\Delta x_{2}<\alpha
\end{eqnarray}
by choosing  input states, and thus one 
obtains (2.17) when one remembers $\epsilon(P)=0$ in the present 
case. To achieve (2.17) it is in fact sufficient to keep 
$\epsilon (x_{1})=\Delta x_{2}$ bounded by a finite constant.

Note that the proof of the conventional relation
\begin{eqnarray}
 \frac{1}{2}\hbar \leq \epsilon(x_{1})\epsilon(P)
\end{eqnarray}
requires that the measurement is unbiased as is explained in 
Appendix, whereas the definition in (2.19) does not satisfy the 
unbiased condition in general. However, the
correlation between $x_{1}$ and $x_{2}$ arising 
from $\epsilon (q)=\epsilon (P)=0$ combined with a crucial {\em 
additional}  assumption 
$\langle (\hat{x}_{2})^{2}\rangle^{1/2}<\alpha$ allows one to 
estimate initial $x_{1}$
to a good accuracy simultaneously with $\epsilon (P)=0$. In 
this sense the above argument
implies the clear evasion of the uncertainty relation for 
$[\hat{x}_{1}, \hat{P}]=i\hbar$. 

In passing, instead of (2.19) one may also consider 
\begin{eqnarray}
\epsilon (x_{1})=\langle (\hat{x}_{2}-c)^{2}\rangle^{1/2}<\alpha
\end{eqnarray}
with a suitable constant $c$, namely, the correlation 
between $x_{1}-c$ and $x_{2}-c$ arising from 
$\epsilon(q)=\epsilon (P)=0$; this notation emphasizes the 
choice of suitable initial states for a given $c$ such that
$\langle (\hat{x}_{2}-c)^{2}\rangle^{1/2}<\alpha$. If one
chooses $c=\langle x_{2}\rangle$, one recovers (2.19) but one
may make a more general choice of $c$. 

The essence of the above analysis is that  the quantity 
$\epsilon (x_{1})=\langle (\hat{x}_{2}-c)^{2}\rangle^{1/2}$ in 
(2.19) or (2.21) is kept 
small by choosing suitable input wave functions, but 
the range covered  by the eigenvalues with 
$\epsilon (x_{1})=\langle (\hat{x}_{2}-c)^{2}\rangle^{1/2}
<\alpha$ is negligible compared to the total space of 
{\em a priori} allowed eigenvalues $-\infty <x_{2}<\infty$. 
Stated differently, one 
can guess the position $x_{1}$ accurately simultaneously with 
$\epsilon(P)=0$ by measuring $q$ precisely, if the 
position $x_{2}$  of the particle 2 happens to be well localized
 around some fixed $c$. The probability of observing such an
event is very small.

 This small probability is  
understood by considering a generic normalized wave function $\Psi(x_{1},x_{2})\simeq \sum_{n} a_{n}\psi_{n}(x_{2},q)$ 
where $\psi_{n}(x_{2},q)$ stand for normalized non-overlapping 
wave packets in $x_{2}$ with a width $\alpha$. The function 
 $\psi_{n}(x_{2},q)$ may have a peak around the precisely measured value of $q$ and the coefficients are 
$|a_{n}|\sim 1/\sqrt{L/\alpha}$ without any additional information about $x_{2}$. The above specific choice of the initial state in (2.21) corresponds to selectively picking up a specific  $\psi_{n}(x_{2},q)$ which contains $x_{2}=c$, but the standard interpretation of  $\Psi(x_{1},x_{2})$ in quantum mechanics predicts that such a probability is negligibly small $|a_{n}|^{2}\sim \alpha/L$ for $L\rightarrow {\rm large}$.

These properties related to (2.13) and (2.21) are the 
general aspects of the clear evasion of the uncertainty 
relation which is recognized by the help of indirect measurements, and the probability for observing  such events which lead to the  clear evasion of the uncertainty relation is negligibly small.  

We also mention that the essential aspects of the evasion of the uncertainty relation in Subsection 2.2 depend 
only on the initial quantum variables and initial quantum 
states, and the analysis depends on the minimal set
 of definitions summarized in Appendix. Our analysis is
insensitive to the specific model of measurements, which in turn suggests that the clear evasion of uncertainty relations discussed here is not an artifact of a specific model of measurements.

No explicit reference to entanglement appears in the above 
analysis. To discuss the implications of entanglement on the 
present problem, one needs a quantitative definition of  entanglement. The entanglement in the context of
quantum information is specified in terms of standard 
deviations~\footnote{ A {\em sufficient} 
condition of entanglement is written as (with a suitable choice of a dimensional constant) 
\begin{eqnarray}
2\hbar> \Delta^{2}(x_{1}-x_{2})+\Delta^{2}(p_{1}+p_{2})\geq 0
\nonumber
\end{eqnarray}
which is also known to be a necessary condition for  Gaussian
two-particle states\cite{simon,englert, fujikawa}. Here 
$\Delta(x_{1}-x_{2})$
and $\Delta(p_{1}+p_{2})$ stand for the standard deviations. 
It is explained in~\cite{englert} that the conventional EPR-type
entanglement is included in this criterion.}.  
We may thus tentatively assume that the precise 
measurements $\epsilon(P)=\epsilon(q)=0$ have been performed for the initial states with $\Delta P=\Delta q=0$, which is possible though there is no strong reason to assume so.  The quantity 
$\epsilon (x_{1})=\langle (\hat{x}_{2})^{2}\rangle^{1/2}$
in (2.19) may then be understood as a measure of correlation between $x_{1}$ and $x_{2}$ arising from $\Delta P=\Delta q=0$ for the initial states.
One may also use the relation $\epsilon(P)=0$ as a resource  
coming from entanglement in the present sense and choose 
$\epsilon(p_{1})=\langle(\hat{P}-\hat{p}_{1})^{2}\rangle^{1/2}
=\langle(\hat{p}_{2})^{2}\rangle^{1/2}$. In this case, however,
 we have $\epsilon(x_{1})\epsilon(p_{1})=
\langle(\hat{x}_{2})^{2}\rangle^{1/2}\langle(\hat{p}_{2})^{2}
\rangle^{1/2}\geq \frac{1}{2}\hbar$ by noting 
$\langle(\hat{x}_{2})^{2}\rangle^{1/2}\geq\Delta x_{2}$, for 
example, and no evasion of the uncertainty relation takes place~\cite{ozawa-pl04}. This argument suggests  
that the connection of the clear evasion of the uncertainty 
relation with the entanglement is not simple and the entanglement by itself is not essential for the evasion of the uncertainty relation.
 
\subsection{Simple model}

To illustrate our analysis  related to (2.17), we 
consider a simple model of two non-interacting particles contained in
 a one-dimensional "box" with a length $L$. We impose a periodic boundary condition, for simplicity. The coordinates are 
then confined in 
\begin{eqnarray}
-\frac{1}{2}L\leq x_{1}, \ x_{2} \leq \frac{1}{2}L
\end{eqnarray}
and the momenta $p_{1}, \ p_{2}$ are given by $p_{1}, \ p_{2}
=2n\pi\hbar/L$ with $n=0,\pm 1,\pm 2,...\ $. Without any 
previous knowledge, the precise measurement of $P$ corresponds 
to the 
distinction of $(n_{1}+n_{2})2\pi\hbar/L$ from 
$(n_{1}+n_{2}\pm 1)2\pi\hbar/L$ for {\em a priori} quite large 
$n_{1}+n_{2}$. The natural estimate of the noise (or error)
in $P$ may be a fraction of $2\pi\hbar/L$ 
\begin{eqnarray}
\epsilon(P)\sim 2\pi\hbar/L,
\end{eqnarray}
and the natural estimate of the noise (or error) in $x_{1}$, 
when one knows only $q=x_{1}-x_{2}$ precisely, is a fraction of 
$L$
\begin{eqnarray}
\epsilon (x_{1})=\langle (\hat{x}_{2}-c)^{2}\rangle^{1/2}\sim L
\end{eqnarray}
for a fixed $c$.
We thus have 
\begin{eqnarray}
\epsilon (x_{1})\epsilon(P) \sim 2\pi\hbar.
\end{eqnarray}
One may of course choose the initial states such 
that  
\begin{eqnarray}
\epsilon (x_{1})=\langle(\hat{x}_{2}-c)^{2}\rangle^{1/2}\leq 
\alpha
\end{eqnarray}
for any small $\alpha$~\cite{ozawa-pl04} and thus 
\begin{eqnarray}
\epsilon (x_{1})\epsilon(P) \sim 2\alpha\pi\hbar/L \rightarrow 0
\end{eqnarray}
for $L\rightarrow \infty$. But the probability
of hitting $\epsilon (x_{1})=\langle(\hat{x}_{2}-c)^{2}
\rangle^{1/2}\leq \alpha$ with a given fixed $c$ in the space of {\em a priori} allowed
 eigenvalues $-\frac{1}{2}L< x_{2}< \frac{1}{2}L$ is negligibly 
small for $L\rightarrow\infty$, {\em without} 
any information such as coming from a preceding direct 
measurement of $x_{2}$. We emphasize that this vanishingly small probability for $\langle(\hat{x}_{2}-c)^{2}\rangle^{1/2}\leq \alpha$ is inevitable and not a result of our incompetence in performing experiments; the probability to find the particle 2 in any finite interval is negligibly small if one only knows that two particles with given $q=x_{1}-x_{2}$ exist in the interval 
$-\infty< x_{1}, \ x_{2}< \infty $ but without any more information. This is confirmed, as already explained in Subsection 2.2, by considering a generic normalized wave function of the form 
$\Psi(x_{1},x_{2})\simeq \sum_{n} a_{n}\psi_{n}(x_{2},q)$ 
where $\psi_{n}(x_{2},q)$ stand for normalized non-overlapping 
wave packets in $x_{2}$ with a width $\alpha$ and $|a_{n}|\sim 1/\sqrt{L/\alpha}$. 

The correlation between 
$x_{1}$ and $x_{2}$ (or between $x_{1}-c$ and $x_{2}-c$) arising
 from $\epsilon(q)=0$ combined with  a 
restriction to a measure-zero set of initial states with 
$\langle(\hat{x}_{2}-c)^{2}\rangle^{1/2}\leq \alpha$ leads to 
the clear evasion of the uncertainty relation (2.17) but with
a negligibly small probability.

\subsection{Additional comment on the diffraction}

We here comment on a treatment of the diffraction process in Subsection 2.1 in the manner of the relations (2.12) and (2.13). 

We consider a combined system of the particle 
(particle 1 with mass $m_{1}$) and the idealized center of the 
slit (particle 2 with mass $m_{2}$) in Fig.1  as a quantum mechanical object~\footnote{Our basic assumption here is that the center of 
the slit is described by an idealized quantum mechanical object 
with a very specific scattering power.}:  We define 
the phase space variables by 
\begin{eqnarray}
&&P_{x}=p_{1x}+p_{2x}, \ \ \ \ \ P_{y}=p_{1y}+p_{2y},
\nonumber\\
&&q_{x}=L=x_{1}-x_{2}, \ \ \ \ \ q_{y}=y=y_{1}-y_{2},
\end{eqnarray}
with conjugate variables
\begin{eqnarray}
&&Q_{x}=(m_{1}x_{1}+m_{2}x_{2})/(m_{1}+m_{2}),\ \ \ \ \ 
Q_{y}=(m_{1}y_{1}+m_{2}y_{2})/(m_{1}+m_{2}), \nonumber\\
&&\tilde{p}_{x}=\frac{m_{2}}{m_{1}+m_{2}}p_{1x}
-\frac{m_{1}}{m_{1}+m_{2}}p_{2x}, \ \ \ \ \ 
\tilde{p}_{y}=\frac{m_{2}}{m_{1}+m_{2}}p_{1y}-\frac{m_{1}}{m_{1}+m_{2}}
p_{2y},
\end{eqnarray}
by treating the problem as a {\em two-dimensional} problem for simplicity with 
$m_{2}\gg m_{1}$.

We start with the particle 1 with $(p_{1x},p_{1y})=(p,0)$ produced  by a particle generator placed in the far left in Fig.1 and 
the slit with $(p_{2x},p_{2y})=(0,0)$. We suppose the steady state with a fixed number of particles passing through the slit per unit time, such as a single particle per second. Assuming the elastic scattering at the slit, $(P_{x},P_{y})=(p,0)$ is preserved, and the magnitude of the relative momentum $\sqrt{\tilde{p}_{x}^{2}+\tilde{p}_{y}^{2}}$ stays constant  since the contribution of the total momentum 
to the total conserved energy is negligible due to the huge mass
$m_{2}$. This is the initial state to analyze the uncertainty relation.

To describe the above quantum state, the position of the slit is crucial and thus we assume the precise measurement of $y_{2}$ and $x_{2}$ with $\epsilon(y_{2})=\epsilon(x_{2})=0$. Then apparently $\eta(p_{1y})=\eta(p_{1x})=0$. This setting may be regarded as corresponding to the fact that 
$p_{y}$  ($=p_{1y}$ in the present notation) in (2.1) and also $p_{1x}$  are not directly measured and $\delta p_{y}\rightarrow 0$ (and also $\delta p_{x}\rightarrow 0$) for $L\rightarrow \infty$ independently of any fixed $\delta y$ and $\delta L$ in (2.2). 
One may then regard the precisely measured apparatus variable $\hat{\mu}_{Y2}$ of 
$y_{2}$ {\em plus} a constant $c$ as an indirect measurement 
apparatus  of $y_{1}$ (and also the precisely measured apparatus variable $\hat{\mu}_{X2}$ of $x_{2}$ plus a constant $L$ as an indirect measurement 
apparatus  of $x_{1}$)~\cite{ozawa-pl03}, and thus the measurement errors 
\begin{eqnarray}
\epsilon(y_{1})
&=&\langle(\hat{y}_{2}+c-\hat{y}_{1})^{2}\rangle^{1/2}
=\langle(\hat{q}_{y}-c)^{2}\rangle^{1/2},\nonumber\\
\epsilon(x_{1})
&=&\langle(\hat{x}_{2}+L-\hat{x}_{1})^{2}\rangle^{1/2}
=\langle(\hat{q}_{x}-L)^{2}\rangle^{1/2}.
\end{eqnarray}
One can then evade the uncertainty relation for 
$[y_{1},p_{1y}]=i\hbar$ (and also for $[x_{1},p_{1x}]=i\hbar$)
in the sense
\begin{eqnarray}
\eta(p_{1y})\epsilon(y_{1})&=&\eta(p_{1y})
\langle(\hat{q}_{y}-c)^{2}\rangle^{1/2}=0, \nonumber\\
\eta(p_{1x})\epsilon(x_{1})&=&\eta(p_{1x})
\langle(\hat{q}_{x}-L)^{2}\rangle^{1/2}=0,
\end{eqnarray}
since $\eta(p_{1y})=\eta(p_{1x})=0$ and one can choose initial states for any fixed $c$ and $L$ such that  
\begin{eqnarray}
\langle(\hat{q}_{y}-c)^{2}\rangle^{1/2}\leq \alpha, \ \ \ 
\langle(\hat{q}_{x}-L)^{2}\rangle^{1/2}\leq \alpha
\end{eqnarray}
for arbitrarily small $\alpha$ by following the argument in~\cite{ozawa-pl03}.

One may recognize that we assumed the measurement of $y ({\rm or} \ q_{y})$ within an interval $\delta y$ in (2.1) and (2.2), whereas in the present case this measurement is replaced by the specification of the constants $c$ and $L$ contained in the indirect measurements and the specification of states which satisfy (2.32). The analysis of the uncertainty relation depends on 
$\delta p_{y}=p(\delta y/L)$ and $\delta y$ in (2.3) and on $\eta(p_{1y})=0$ and $\alpha$ in the present case, and  the analysis of the uncertainty relation itself is not sensitive to the actual value of $y$ or $c$.
We consider that the present setting is closer to the actual situation described in Fig. 1. In particular, the estimate of the error  $\delta p_{y}=p(\delta y/L)$ in (2.2) is based on an expected error when one specifies $y$ with an accuracy $\delta y$ without taking into account the possible disturbance in $p_{y}$ which is caused by any experimental specification of $y$.

The normalized wave function for the configuration in  Fig.1 may be 
described by a superposition of wave functions at $q_{x}=L$ (by concentrating on the freedom in the y-direction) such as 
\begin{eqnarray}
\Psi(y_{1},y_{2})\simeq \sum_{n=1}^{N}a_{n}\psi_{n}(q_{y},y_{2})
\end{eqnarray}
where each $\psi_{n}(q_{y},y_{2})$ gives a normalized 
non-overlapping wave packet in $q_{y}$ with a size 
$\alpha(=\delta y)$, and 
\begin{eqnarray}
|a_{n}|\sim \frac{1}{\sqrt{N}}\sim \frac{1}{\sqrt{\tilde{L}/\alpha}}
\end{eqnarray}
without any additional information about $q_{y}$. Here $\tilde{L}$ is a length parameter which characterizes the spread of the diffracted particle at $q_{x}=L$, and $\tilde{L}
\rightarrow\infty$ for $L\rightarrow\infty$. 
The above special choice of the initial state in (2.32) then 
corresponds to selectively picking up a special $\psi_{n}(q_{y},y_{2})$ which is centered at $q_{y}\simeq c$, but the standard interpretation of $\Psi(y_{1},y_{2})$ in quantum mechanics predicts that  such a probability is vanishingly small $|a_{n}|^{2}\sim \alpha/\tilde{L}\rightarrow 0$ for $L\rightarrow\infty$. This is equivalent to choosing a 
vanishingly narrow phase space. One may also use the outgoing wave in the radial direction such as (2.9) in this analysis.

 If one fixes $y$ or $c$ together with $L$ by any means, one can estimate the value of $p_{y}$ by using the geometrical information (2.1). But we consider that the estimate of the value of $p_{y}$ itself  is not essential in the analysis of the uncertainty relation,
 since the estimate of $p_{y}$ depends on the pre-determined magnitude of the relative momentum $\sqrt{\tilde{p}_{x}^{2}+\tilde{p}_{y}^{2}}$. 
Even in the context of the relations (2.12) and (2.13) in Subsection 2.2, one may slightly change the setting of the problem and perform an analogous analysis. One may assume that a steady flux of the particle with momentum $p$ is supplied by a particle generator placed in the far left and the flux is passing through a box with a length $L$ in one-dimensional space. Then the wave function of the particle is expected to be of the 
form $\psi(x)\sim\frac{1}{\sqrt{L}}\exp[ipx/\hbar]$ inside the box. (The precise measurement of the coordinate of the particle 2 in the example of (2.12) and (2.13) may be regarded as a specification of the origin of the coordinate, since the particle 2 has no interaction with the particle 1 and the determination of the position  $x_{2}$ provides the relative origin to measure $x_{1}$.) If one assumes the observation of the particle with the well-defined momentum $p$ in a small interval $\delta x$, it leads to a clear evasion of the uncertainty relation but such a probability is very small $|\psi(x)|^{2}\delta x\sim \delta x/L$.  After an actual observation of the particle location with an accuracy $\delta x$, it is described by a wave packet of the size $\delta x$. 
A state described by such a wave packet is then realized with a unit probability due to the reduction of the state; if one assumes the simultaneous observation of the specific well-defined momentum $p$, it then leads to a clear evasion of the uncertainty relation. But this time the probability of observing a state with the well-defined momentum $p$ inside the wave packet is very small, as is confirmed by performing a Fourier analysis. We thus conclude that  the probability of the clear evasion of the uncertainty relation is very small in any case.   
In contrast, if one wants to observe a particle described by a plane wave with nearly unit probability, for example, one needs to cover the range of the order $L$ in $x$ space and then no clear evasion of the uncertainty relation occurs.

\subsection{Preparation of initial states}

In all the examples of the clear evasion of the uncertainty 
relation we discussed so far, we have explained that the 
initial states are assumed to be confined in a subspace of 
the total {\em a priori} allowed phase space, which has a very small measure. This means that 
the probability of observing such events is very small.
On the other hand, if one prepares the initial states by
 some means and treats the preparation
 as a part of measurement, it is
 then shown that the uncertainty relation is not clearly 
evaded. This 
feature appears to be quite general~\cite{ballentine}. 

Let us repeat the analysis of the case associated with (2.13).
If one does not know anything about the value of $q=x_{1}-x_{2}$
by any previous measurement, the value of $q$ can be   
very small as well as very large. One may then assume that
the value of $q$ is confined in between $q$ and $q+\alpha$,
though no reasons to assume so.
Combined with the precise measurement of $x_{2}$, one can then estimate $x_{1}$  within the accuracy $\alpha$
which one can choose to be arbitrarily small without disturbing 
$\eta(p_{1})=0$. However, the probability for meeting such a lucky situation for the actual initial states, which specify only the value of $x_{2}$ precisely, is 
\begin{eqnarray}
{\rm Probability}=|\psi(q)|^{2}\alpha \sim \frac{\alpha}{L}
\end{eqnarray}
as is explained in (2.16) by using a different notation. See also (2.10) and (2.34). The probability thus becomes vanishingly small for 
$L\rightarrow \infty$, as we discussed already.

On the other hand, one may attempt to prepare the sample
, which has a spread in $q$ between $q$ and $q+\alpha$,
by some preceding measurements. In this case, the final state of
the preceding measurements now becomes the initial state
in (2.12) and (2.13). The normalization factor $1/\sqrt{L}$ does
 not appear in the initial state thus defined due to 
the ``reduction'' of the wave function,  but one 
{\em inevitably} 
induces the disturbance of the order of $\hbar/\alpha$ in the 
conjugate relative momentum $p=(p_{1}-p_{2})/2$ and thus in $p_{1}$ 
also, namely, $\eta(p_{1})\sim \frac{\hbar}{\alpha}$. (Note that
the preparation has a meaning only when it takes place with 
a sizable probability.) One may 
then perform the precise measurement of $x_{2}$ as described 
in (2.12) and (2.13). One can 
determine the position $x_{1}$ by the knowledge of $q$, which is
localized  between 
$q$ and $q+\alpha$, and the precisely measured $x_{2}$. In this 
case we have
\begin{eqnarray}
\eta(p_{1})\delta x_{1}\sim \frac{\hbar}{\alpha}\times \alpha
\sim \hbar
\end{eqnarray}
where $\delta x_{1}$ is an error in the estimate of $x_{1}$, and we have no clear evasion of the uncertainty relation for the combined process of  preparation and measurement.  

Similarly, in the analysis of (2.17) a preceding measurement of
  $x_{2}$ with an accuracy $\delta x_{2}$ induces the disturbance
$\sim\hbar/\delta x_{2}$ in the variable $p_{2}$ and thus in $P$ also. (Note again that
the preparation has a meaning only when it takes place with 
a sizable probability.) We 
thus have 
\begin{eqnarray}
\epsilon(x_{1})\delta P \sim \Delta x_{2}\delta P
\sim \hbar
\end{eqnarray}
 in (2.17), where we used (2.19) with $\Delta x_{2}\sim \delta x_{2}$ for a system of which $x_{2}$ is determined with an accuracy $\delta x_{2}$, and thus no clear 
evasion of the uncertainty relation. Note that we perform the 
measurement immediately after the preparation~\footnote{If one does not perform the measurement immediately after the preparation, the system is generally settled to a quite different state as is seen in the case of diffraction process in Fig.1 where the initial fluctuation in $p_{y}$ immediately after the passage of the slit is eventually absorbed into the effect of diffraction.} , and thus the uncertainty $\delta P$ in $P$, which is induced by the entire measurement
process, is a sum of $\sim\hbar/\delta x_{2}$ and  $\epsilon(P)=0$ in the present case.

\section{Uncertainty relation in a two-spin system with EPR entanglement}

We here comment on  the uncertainty relation for a two-spin 
system $\vec{s}_{1}$ and $\vec{s}_{2}$ with the EPR 
entanglement. To be precise, our analysis is valid for any two-spin system which is in a general inseparable state since our analysis holds regardless of the spatial separation of two spins as long as the interaction between the spins is ignored.  Physically it is most interesting to regard our system  as corresponding to a two-spin system spatially far apart as is discussed in \cite{bohm}.

We analyze the uncertainty relation associated with the 
generic non-Abelian structure of the angular 
momentum~\cite{robertson}
\begin{eqnarray}
[\hat{J}_{k}, \hat{J}_{l}]=i\hbar \epsilon^{klm}\hat{J}_{m}.
\end{eqnarray}
One may simultaneously measure spin components $s_{1z}$ 
and $s_{2y}$ precisely since these variables are commuting.
We thus have
\begin{eqnarray}
\epsilon(s_{1z})=0, \ \ \ \ \ \epsilon(s_{2y})=0.
\end{eqnarray}
After the simultaneous precise measurements, we will have, 
for example,
\begin{eqnarray}
\psi_{\rm final}=u_{-}(1)v_{+}(2)
\end{eqnarray}
which is a superposition of the total spin $S=0$ and $S=1$. 
Here we defined $\hat{s}_{1y}v_{\pm}(1)=\pm\frac{1}{2}\hbar 
v_{\pm}(1)$ 
and $\hat{s}_{2y}v_{\pm}(2)=\pm\frac{1}{2}\hbar v_{\pm}(2)$, and 
also $\hat{s}_{1z}u_{\pm}(1)=\pm\frac{1}{2}\hbar u_{\pm}(1)$
and $\hat{s}_{2z}u_{\pm}(2)=\pm\frac{1}{2}\hbar u_{\pm}(2)$.
This $\psi_{\rm final}$ is the information we obtain by 
simultaneous precise measurements $\epsilon(s_{1z})=0$ and 
$\epsilon(s_{2y})=0$ without assuming any entanglement in the 
initial state.
The basic question is then if one obtains any extra information 
about the spin $s_{1y}$ or spin $s_{2z}$ in the initial state 
by assuming the entanglement specified by $S=0$, for example.

By assuming the entanglement specified by $S=0$, for example, we have
\begin{eqnarray}
\hat{S}_{y}|S=0, S_{y}=0\rangle=
(\hat{s}_{1y}+\hat{s}_{2y})
|S=0, S_{y}=0\rangle=0
\end{eqnarray}
and 
\begin{eqnarray}
&&\langle S=0, S_{y}=0|\hat{s}_{1y}\hat{s}_{2y}
|S=0, S_{y}=0\rangle\nonumber\\
&&=-\frac{1}{4}\hbar^{2}\\
&&\neq
\langle S=0, S_{y}=0|\hat{s}_{1y}|S=0, S_{y}=0\rangle
\langle S=0, S_{y}=0|\hat{s}_{2y}|S=0, S_{y}=0\rangle=0\nonumber
\end{eqnarray}
which follows from 
$\langle S=0, S_{y}=0|\hat{S}^{2}_{y}|S=0, S_{y}=0\rangle
=0$.
It thus appears that one can estimate the precise value of 
$s_{1y}$ indirectly by means of entanglement from the precisely 
measured $s_{2y}$ with $\epsilon(s_{2y})=0$, simultaneously 
with the precise measurement of $s_{1z}$ with 
$\epsilon(s_{1z})=0$. This would suggest the simultaneous 
precise determination of $s_{1y}$ and $s_{1z}$, and thus the 
uncertainty relation for $s_{1y}$ and $s_{1z}$ appears to be 
evaded.

This argument however does not quite work if one remembers that the
symmetry consideration with respect to $1\leftrightarrow 2$ implies the
precise simultaneous estimate of $s_{2y}$ and $s_{2z}$
also, and thus the  precise simultaneous information of $s_{1y}$,
$s_{1z}$, $s_{2y}$ and $s_{2z}$.
 One first recalls the relation for the state with $S=0$
\begin{eqnarray}
\psi_{initial}(0)&=&|S=0, S_{y}=0\rangle
\nonumber\\
&=&\frac{i}{\sqrt{2}}[v_{+}(1)v_{-}(2)-v_{-}(1)v_{+}(2)]
\nonumber\\
&=&\frac{1}{\sqrt{2}}[u_{+}(1)u_{-}(2)-u_{-}(1)u_{+}(2)]
\nonumber\\
&=&|S=0, S_{z}=0\rangle
\end{eqnarray}
where we used $u_{+}=(v_{+}+ v_{-})/\sqrt{2}$ and 
$u_{-}=(v_{+}- v_{-})/(\sqrt{2}i)$.

After the precise measurement of $s_{2y}=\frac{1}{2}\hbar$, for
example, one may expect the wave function to be
\begin{eqnarray}
\psi_{\rm final}=v_{-}(1)v_{+}(2)
\end{eqnarray}
which is the basis why we guess $s_{1y}=-\frac{1}{2}\hbar$ and 
thus obtain extra information.
Similarly, after the precise measurement of 
$s_{1z}=-\frac{1}{2}\hbar$, for
example, one may expect the wave function to be
\begin{eqnarray}
\psi_{\rm final}=u_{-}(1)u_{+}(2)
\end{eqnarray}
which is the basis why we guess $s_{2z}=\frac{1}{2}\hbar$ and 
obtain extra information.
But  
\begin{eqnarray}
v_{-}(1)v_{+}(2)\neq u_{-}(1)u_{+}(2)
\end{eqnarray}
since both of $v_{-}(1)v_{+}(2)$ and $u_{-}(1)u_{+}(2)$ are the 
superposition of $S=0$ and $S=1$, and thus a rotation of 90 degrees
around the $x$ axis does not leave the wave function invariant.
This relation (3.9) shows that the simultaneous accurate 
information about $s_{1z}$ and $s_{2z}$ implied by the 
right-hand side and the simultaneous 
accurate information about $s_{1y}$ and $s_{2y}$ implied by the 
left-hand side cannot coexist and thus cannot be realized.
Thus we cannot have the  precise simultaneous information of 
$s_{1y}$, $s_{1z}$, $s_{2y}$ and $s_{2z}$, and consequently 
no clear evasion of the uncertainty relation.

Another possibility one may examine is to precisely measure 
only $s_{1z}$ and no more measurement. This is analogous to 
(2.12). One may then naively expect
$\eta(s_{2y})=0$. Assuming entanglement with $S=0$ in the 
initial state, one may guess $s_{2z}$ from the measured value 
$s_{1z}$. The information about $s_{2z}$ without disturbing 
$s_{2y}$ with $\eta(s_{2y})=0$ would suggest an evasion of 
the uncertainty relation. This possibility can be examined 
explicitly by noting that the state after the measurement of 
$s_{1z}$ is given by 
\begin{eqnarray}
\psi_{\rm final}=u_{+}(1)u_{-}(2)=\frac{1}{2i}
(v_{+}(1)+ v_{-}(1))(v_{+}(2)- v_{-}(2))   
\end{eqnarray}
which may be compared to the initial state, namely, the first expression in (3.6)
\begin{eqnarray}
\psi_{\rm initial}
=\frac{i}{\sqrt{2}}[v_{+}(1)v_{-}(2)-v_{-}(1)v_{+}(2)].
\end{eqnarray}
From this comparison, one first concludes that $s_{1y}$
is influenced by the precise measurement of $s_{1z}$, but 
from the symmetry under $1\leftrightarrow 2$ in these 
expressions one also
concludes that $s_{2y}$ cannot stay undisturbed by 
the precise measurement of $s_{1z}$. To the extent that 
$\eta(s_{1y})\neq0$ under the precise measurement of $s_{1z}$,
one concludes that $\eta(s_{2y})\neq0$ under the precise 
measurement of $s_{1z}$ and thus no clear evasion of the  
uncertainty relation in the present example. 

This example shows that a clear evasion of the uncertainty 
relation is not realized in the present system with a finite 
number of discrete degrees of freedom even if one uses the indirect determination on the basis of entanglement. This example  also provides a further indication that the 
entanglement by itself is not responsible for the clear
evasion of the uncertainty relation.

\section{Discussion and conclusion}

Motivated by the examples of the clear evasion of the uncertainty relation in the diffraction process by 
Ballentine~\cite{ballentine} and in the two simple gedanken experiments by Ozawa~\cite{ozawa-pl03,ozawa-pl04}, we analyzed
the basic mechanism of the clear evasion of the uncertainty relation. We have shown that the clear evasion of the
uncertainty relation is 
realized in a tiny sector of the phase space with a vanishingly small measure. Namely, the clear evasion of the uncertainty relation is possible but such a probability is very small. This analysis
is consistent with the fact that no clear evasion of the uncertainty relation in experiments has been reported. Our emphasis here is that the uncertainty relation including the Kennard's relation is of probability theoretical nature, as are all the laws in quantum mechanics. We take the uncertainty relation  as a consequence of quantum mechanics rather than as a principle.

We have also argued that, if one considers 
the preparation of the initial states and then the 
analysis of the uncertainty relation later, the combined total process does not lead to the clear evasion of the uncertainty relation. 

In the course of our analysis, we indicated that the
entanglement by itself is not essential for the clear evasion of the 
uncertainty relation. In retrospect, this is quite natural 
since the entanglement or inseparability is the property of state vectors specified by means of the Kennard's uncertainty 
relation~\cite{simon, englert, fujikawa}.   

We also think the following fact is worth mentioning: The
essential aspects of our analysis of the uncertainty relation depend on only the initial quantum 
variables and initial quantum states, and thus the analysis is 
rather insensitive to the detailed model of  measurements. 
If the analysis should be  very sensitive to a specific 
model of measurements, one would naturally ask what 
will happen if one adopts another model of measurements.

Finally,  a remaining more fundamental and difficult 
issue is an analysis of quantitative conditions for the 
transitional region where the evasion of the uncertainty relation starts visible, instead of the present analysis which is limited to the case of the clear evasion of the uncertainty relation, and it is left for a future investigation. If such an
analysis is performed, an experimental test of the idea presented here will become possible.

\appendix

\section{Summary of some basic quantities}
We work in the Heisenberg picture and define the coordinate and momentum observables of a system with one degree 
of freedom before the measurement 
\begin{eqnarray}
(\hat{x}_{i}, \ \ \ \hat{p}_{i})
\end{eqnarray}
and those observables after the measurement 
\begin{eqnarray}
(\hat{x}_{f}, \ \ \ \hat{p}_{f}).
\end{eqnarray}
We also define the observables of the measurement apparatus
after the measurement
\begin{eqnarray}
(\hat{\mu}_{Xf}, \ \ \ \hat{\mu}_{Pf})
\end{eqnarray}
which are mutually {\em commuting} 
$[\hat{\mu}_{Xf}, \hat{\mu}_{Pf}]=0$ and also with 
the final state variables 
$[\hat{\mu}_{Xf},\hat{x}_{f}]=[\hat{\mu}_{Xf},\hat{p}_{f}]=0$
and $[\hat{\mu}_{Pf},\hat{x}_{f}]
=[\hat{\mu}_{Pf},\hat{p}_{f}]=0$.
We then define the measurement error operators by~\cite{appleby,
ozawa-pl03, ozawa-pl04}
\begin{eqnarray}
&&\hat{\epsilon}_{x}=\hat{\mu}_{Xf}-\hat{x}_{i},
\nonumber\\
&&\hat{\epsilon}_{p}=\hat{\mu}_{Pf}-\hat{p}_{i} 
\end{eqnarray}
and the disturbance operators as the difference between initial
 and final operators  
\begin{eqnarray}
&&\hat{\delta}_{x}=\hat{x}_{f}-\hat{x}_{i},
\nonumber\\
&&\hat{\delta}_{p}=\hat{p}_{f}-\hat{p}_{i}. 
\end{eqnarray}
Finally, we define the root-mean-square quantities 
\begin{eqnarray}
&&\epsilon(x)=\langle \psi\otimes \phi_{ap}| 
\hat{\epsilon}_{x}^{2}
|\psi\otimes \phi_{ap}\rangle^{1/2}
\nonumber\\
&&\epsilon(p)=\langle \psi\otimes \phi_{ap}
|\hat{\epsilon}_{p}^{2}
|\psi\otimes \phi_{ap}\rangle^{1/2}
\nonumber\\
&&\eta(x)=\langle \psi\otimes \phi_{ap}
|\hat{\delta}_{x}^{2}
|\psi\otimes \phi_{ap}\rangle^{1/2}
\nonumber\\
&&\eta(p) =\langle \psi\otimes \phi_{ap}
|\hat{\delta}_{p}^{2}
|\psi\otimes \phi_{ap}\rangle^{1/2}
\end{eqnarray}
with the Heisenberg picture state of the form
$|\psi\otimes \phi_{ap}\rangle$ where $\psi$ and $\phi_{ap}$ 
stand for the 
initial states of the system and apparatus, respectively.

The uncertainty relations we adopt in our paper 
are then written in the form 
\begin{eqnarray}
\epsilon(x)\epsilon(p)\geq \frac{1}{2}\hbar
\end{eqnarray}
or in the form 
\begin{eqnarray}
\epsilon(x)\eta(p)\geq \frac{1}{2}\hbar, \ \ \ \ \  
\epsilon(p)\eta(x)\geq \frac{1}{2}\hbar
\end{eqnarray}
where the relation in (A.7) emphasizes the simultaneous measurements of two conjugate quantities, while the relations 
in (A.8) emphasize the measurement-disturbance relations~\cite{appleby,
ozawa-pl03, ozawa-pl04}.
The first relation (A.7) is proved by assuming that the 
measurement is unbiased in the sense
\begin{eqnarray}
&&\langle \psi\otimes \phi_{ap}| \hat{\epsilon}_{x}
|\psi\otimes \phi_{ap}\rangle=0,\nonumber\\
&&\langle \psi\otimes \phi_{ap}| \hat{\epsilon}_{p}
|\psi\otimes \phi_{ap}\rangle=0
\end{eqnarray}
for {\em all} $\psi$~\cite{appleby}. The quantities 
$\epsilon(x)$ and
  $\epsilon(p)$ may be understood as errors involved in the 
measurement process. It should however be noted that the essential 
aspects of the clear evasion of the uncertainty relation in the body of the present paper are understood in the more conventional framework as presented in~\cite{ballentine}.  

The standard deviations are defined for any initial state by
\begin{eqnarray}
&&\Delta x=\langle \psi|( \hat{x} - 
\langle \psi|\hat{x}|\psi\rangle)^{2}|\psi\rangle^{1/2}
\nonumber\\
&&\Delta p=\langle \psi|( \hat{p} - 
\langle \psi|\hat{p}|\psi\rangle)^{2}|\psi\rangle^{1/2}
\end{eqnarray}
and the Kennard's relation
\begin{eqnarray}
\Delta x\Delta p\geq \frac{1}{2}\hbar
\end{eqnarray}
always holds precisely. $\Delta x$ and $\Delta p$ may be  
understood as intrinsic uncertainties in the initial state.


\begin{thebibliography}{99}
\bibitem{dirac}
P.A.M. Dirac, {\em Principles of Quantum Mechanics} (Oxford 
University Press, Oxford, 1958).
\bibitem{heisenberg}
W. Heisenberg, Z. Phys. {\bf 43} (1927) 172.
\bibitem{kennard}
E.H. Kennard, Z. Phys. {\bf 44} (1927) 326.
\bibitem{robertson}
H.P. Robertson, Phys. Rev. {\bf 34} (1929) 163.
\bibitem{epr}
A. Einstein, B. Podolsky and N. Rosen, Phys. Rev. {\bf 47} (1935)
777.
\bibitem{neumann}
J. von Neumann, {\em Mathematical Foundations of Quantum 
Mechanics} (Princeton University Press, Princeton, 1955).
\bibitem{margenau}
H. Margenau, Phil. Sci. {\bf 25} (1958) 23.\\
E. Purgovecki, Found. Phys. {\bf 3} (1973) 3; J. Math. Phys.
{\bf 17} (1976) 1673.
\bibitem{arthurs}
E. Arthurs and J.L. Kelly Jr., Bell System Technical Journal,
{\bf 44} (1965) 725.   
\bibitem{ballentine}
L.E. Ballentine, Rev. Mod. Phys. {\bf 42} (1970) 358.
\bibitem{braginsky}
V.B. Braginsky and F. Ya. Khalili, {\em Quantum Measurement},
(Cambridge University Press, Cambridge, 1992).
\bibitem{appleby}
D.M. Appleby, Int. J. Theor. 
Phys. {\bf 37} (1998) 1491.\\
D.M. Appleby, Jour. Phys. A{\bf 31} (1998) 6419,
 and references cited therein.
\bibitem{appleby2}
D.M. Appleby, Int. J. Theor. 
Phys.
{\bf 37} (1998) 2557.
\bibitem{busch}
P. Busch and P.J. Lahti, Found. Phys. {\bf 19} (1989) 633.
\bibitem{hilgevoord}
J. Hilgevoord and J. Uffink, in {\em Sixty Two Years of 
Uncertainty}, edited by A.I. Miller, (Plenum Press, New York, 
1990).
\bibitem{uffink}
J. Uffink, Int. J. Theor. Phys. {\bf 33} (1994) 199.
\bibitem{busch2}
P. Busch, M. Grabowski and P.J. Lahti, {\em Operational Quantum 
Physics}, (Springer-Verlag, Berlin, 1995).
\bibitem{ozawa-pl03}
M. Ozawa, Phys. Lett. A{\bf 318} (2003) 21.
\bibitem{ozawa-pl04}
M. Ozawa, Phys. Lett. A{\bf 320} (2004) 367.
\bibitem{simon}
R. Simon, Phys. Rev. Lett. {\bf 84} (2000) 2726.
\bibitem{englert}
B.G. Englert and K. Wodkiewicz, Phys. Rev. A{\bf 65} (2002)
 054303.
\bibitem{fujikawa}
K. Fujikawa, ``On the separability criterion for continuous 
variable systems'', arXiv:0710.5039.
\bibitem{bohm}
D. Bohm, {\em Quantum Mechanics} (Prentice-Hall, Englewood 
Cliffs, 1951).
%\bibitem{peres}
%A. Peres, Phys. Rev. Lett. {\bf 77} (1996) 1413.
%\bibitem{horodecki}
%P. Horodecki, Phys. Lett. A{\bf 232} (1997) 333.
\end{thebibliography}
\end{document}